\documentclass[letterpaper, 10 pt, conference]{ieeeconf}  

\IEEEoverridecommandlockouts                              

\usepackage{graphics} 
\usepackage{epsfig} 
\usepackage{mathptmx} 
\usepackage{times} 
\usepackage{amsmath} 
\usepackage{amssymb}  
\usepackage{multirow} 
\usepackage{hyperref}

\title{\LARGE \bf
A Review on Trajectory Datasets on Advanced Driver Assistance System
}

\author{Hang Zhou, Ke Ma*, Xiaopeng Li
\thanks{*Research supported by the U.S. National Science Foundation under Grant No.2313578. (corresponding author: Ke Ma). Hang Zhou, Ke Ma, and Xiaopeng Li are now with the Department of Civil and Environmental Engineering, Univ. of Wisconsin-Madison, Madison, WI, USA. (e-mails: kma62@wisc.edu).}%
}

\begin{document}

\maketitle
\thispagestyle{empty}
\pagestyle{empty}

\begin{abstract}
This paper presents a comprehensive review of trajectory data of Advanced Driver Assistance System equipped-vehicle, with the aim of precisely model of Autonomous Vehicles (AVs) behavior. This study emphasizes the importance of trajectory data in the development of AV models, especially in car-following scenarios. We introduce and evaluate several datasets: the OpenACC Dataset, the Connected \& Autonomous Transportation Systems Laboratory Open Dataset, the Vanderbilt ACC Dataset, the Central Ohio Dataset, and the Waymo Open Dataset. Each dataset offers unique insights into AV behaviors, yet they share common challenges in terms of data availability, processing, and standardization. After a series of data cleaning, outlier removal and statistical analysis, this paper transforms datasets of varied formats into a uniform standard, thereby improving their applicability for modeling AV car-following behavior. 
Key contributions of this study include: 1. the transformation of all datasets into a unified standard format, enhancing their utility for broad research applications; 2. a comparative analysis of these datasets, highlighting their distinct characteristics and implications for car-following model development; 3. the provision of guidelines for future data collection projects, along with the open-source release of all processed data and code for use by the research community. 

\end{abstract}

\section{INTRODUCTION}

The rapid development of Autonomous Vehicles (AVs) necessitates a more profound understanding of their behavior to ensure traffic safety and efficiency. This involves developing strong, real-world models that accurately capture the complex behaviors of AVs. Substantial advancements have been made in the development of models to analyze AV behavior in varied scenarios such as car-following, platooning, intersection navigation, and lane-changing maneuvers. These models demand fine-tuned parameters, underpinned by real-world empirical data.

Recently, a surge in autonomous driving datasets like BDD100K \cite{yu2020bdd100k}, Argoverse \cite{chang2019argoverse, wilson2023argoverse}, and Waymo Open Dataset \cite{mei2022waymo} has been observed, with substantial reviews in the literature \cite{liu2024survey, liu2021survey}. Primarily, these datasets concentrate on perception data collected by cameras, LiDAR, rader, Inertial Measurement Units (IMU), and other sensors. Although these perception data can ensure the basic safety conditions and be used to develop simple dynamics of AV, they are not accurate enough for complex behavior model. This oversight constrains the progression of detailed AV-specific traffic flow analysis. 

To address the need for accurate vehicle dynamics behavior modeling in future transportation construction and operation, trajectory data enriched with Global Position System (GPS) information emerges as a critical resource. This data type, superior in accuracy to perception data, is indispensable for developing sophisticated AV dynamics models. Such models require data that encapsulate a wide array of scenarios (like rural road, highway, or intersection), cases (car-following, lane-changing, or merging), and states (low/high speed following, cruise control, or emergency deceleration), with precise measurements of vehicle position and speed. Consequently, leveraging GPS to gather extensive trajectory data presents an economically viable alternative to perception data.

However, the availability of trajectory data is currently quite limited. Compounding this issue is the challenge of processing trajectory data due to inconsistencies in acquisition methods, equipment used (different GPS models, and whether they carry IMUs), vehicles involved (brands, models, and years), road environments, and most importantly, data release standards. These disparities hinder the development of uniform datasets essential for calibrating AV behavioral models, like traditional linear model (will be further explained below), and simulating authentic driving behavior. The existing literature underscores the paucity of datasets concentrated on trajectory data, with the added complication of variability in data formats and collection methodologies, including differences in sampling frequencies. This situation delineates a significant research gap; there is an urgent need for comprehensive review that summarizes and analyzes current trajectory datasets and provides guideline to the future trajectory data collection projects. 

In addressing these challenges, this study undertakes a comprehensive review of existing open datasets that include trajectories data, particularly from vehicles equipped with Advanced Driver Assistance System (ADAS). We focus on the dataset with car-following trajectories, since it is the most important model in the traffic follow analysis. Our objective is to transfer these datasets into a uniform, user-friendly format. This proposed standard can serve as a benchmark for future trajectory datasets. Our methodology encompasses the extraction of main features that related to the car-following behavior, data cleaning and the removal of outliers. Our research also includes a statistical analysis and the calibration of these datasets. Post-acceptance of this paper, we intend to make all our methods, codes, and refined datasets accessible to the wider community, thereby facilitating seamless integration into our proposed framework for subsequent research endeavors. Our key contributions are as follows:
\begin{itemize}
        \item We summarize the existing open trajectory datasets on ADAS, which provide a comprehensive optional data list for the traffic flow analysis especially for the development of the car-following model. 
        \item We transform three datasets, i.e., the OpenACC Dataset, the Connected \& Autonomous Transportation Systems (CATS) Laboratory Open Dataset, and the Vanderbilt ACC Dataset, into a uniform and standardized format. This standard format is expected to provide a guideline to collect and publish trajectory data for future studies.
        \item Then, an initial data cleaning is conducted to fill in missing data and remove outlier and unstable data. The improved data has a higher qualify for car-following model development.
        \item In the experiments, we compare the statistical analysis and the calibration results of a linear car-following model to three selected datasets. Results show that the CATS Laboratory Open Dataset and the Vanderbilt ACC Dataset are suitable for middle speed model, while the OpenACC Dataset is better for low speed model.
\end{itemize}

The remainder of this paper is organized as follows: Section \ref{sec:review} reviews related works on existing AV trajectory datasets. Section \ref{sec:method} introduces our data processing method. Section \ref{sec:exp} describes the calibration results. Finally, conclusions are presented in Section \ref{sec:con}.	

\section{RELATED WORKS}
\label{sec:review}

The availability of high-quality and publicly accessible datasets is crucial for the development and validation of novel AV control models. Due to the combined efforts of research institutions and the industry, a significant number of datasets have been developed over recent decades. These datasets have been thoroughly reviewed and evaluated in academic literature, including works by entities like the IEEE Emerging Transportation Technology Testing (ET3) Committee. This section primarily focuses on reviewing datasets that feature extended car-following trajectories, specifically those exceeding 15 seconds in duration. Noteworthy datasets containing shorter or partial trajectories, such as the Argoverse Motion Dataset \cite{chang2019argoverse, wilson2023argoverse}, are excluded from our review. Table \ref{tab:summary} presents a compilation of existing representative datasets.

\begin{table*}[h]
        \caption{AUTONOMOUS DRIVING DATASETS with trajectory data}
        \begin{center}
        \begin{tabular}{p{8em}p{10em}p{10em}p{20em}p{6em}}
        \hline
        Name  & Main features & Size & Format & Roadway type \\
        \hline
        Vanderbilt ACC Dataset \cite{wang2019estimating}  & time, speed, spacing, the follower's acceleration & 15 minutes continuous trajectory data. & All data stored in one CSV file. Each row contains one frame data for two vehicles. & Freeway \\
        OpenACC Dataset \cite{makridis2021openacc} & time, speed, latitude, longitude, altitude, spacing & 28 vehicles with more than 3000 km trajectory data.  & Including seven folders, each folder contains several CSV files. Each row contains one frame data for multiple vehicles. & Highway, Urban, Test track \\
        Central Ohio Dataset \cite{OhioOneVehicles,OhioTwoVehicles} & time, headway, position (x,y), speed, acceleration, lane\_id & 120 hours trajectory data, with intotal 4169447 rows. & All data stored in one CSV file. Each row contains one frame data for multiple vehicles. & Highway, Freeway \\
        CATS Laboratory Open Dataset \cite{shi2021empirical} & time, latitude, longitude, speed &  Two 25 minutes continuous trajectory data.  & Each vehicle's data is stored in one XLSX file. Each sheet contain one trajectory. & Urban \\
        Waymo Open Dataset processed by Hu et al. \cite{hu2022processing} & time, position, speed, acceleration, jerk & 470 AV-related car-following pairs with about 20 seconds each. & All data stored in one CSV file. Each row contains one frame data for two vehicles & Downtown, Suburban \\
        \hline
        \end{tabular}%
        \label{tab:summary}%
\end{center}       
\end{table*}%

\subsection{OpenACC Dataset}

The OpenACC Dataset \cite{makridis2021openacc}, collated by the European Commission, offers an insightful perspective on the behavior of commercial Adaptive Cruise Control (ACC) systems under diverse driving conditions. This dataset encompasses data from 28 vehicles, including internal combustion engine (ICE) vehicles, electric vehicles, and hybrids, with 22 of these vehicles being outfitted with cutting-edge commercial ACC systems. To gather comprehensive data, the collection team embarked on four experimental campaigns, aimed at rigorously testing vehicle dynamics in real-world conditions, as well as the functionality of ACC systems and car-following patterns. In the OpenACC dataset, data from these four campaigns are methodically organized into corresponding folders: Cherasco, Vicolungo, AstaZero, and ZalaZone. Each of these folders contains several CSV files, with each file documenting the trajectory of a vehicle platoon. In total, the dataset encompasses over 3000 kilometers of trajectory data, captured at a resolution of 10 Hz.

\subsection{Connected \& Autonomous Transportation Systems (CATS) Laboratory Open Dataset}

\begin{figure}
\centering
\includegraphics[width=1.0\linewidth]{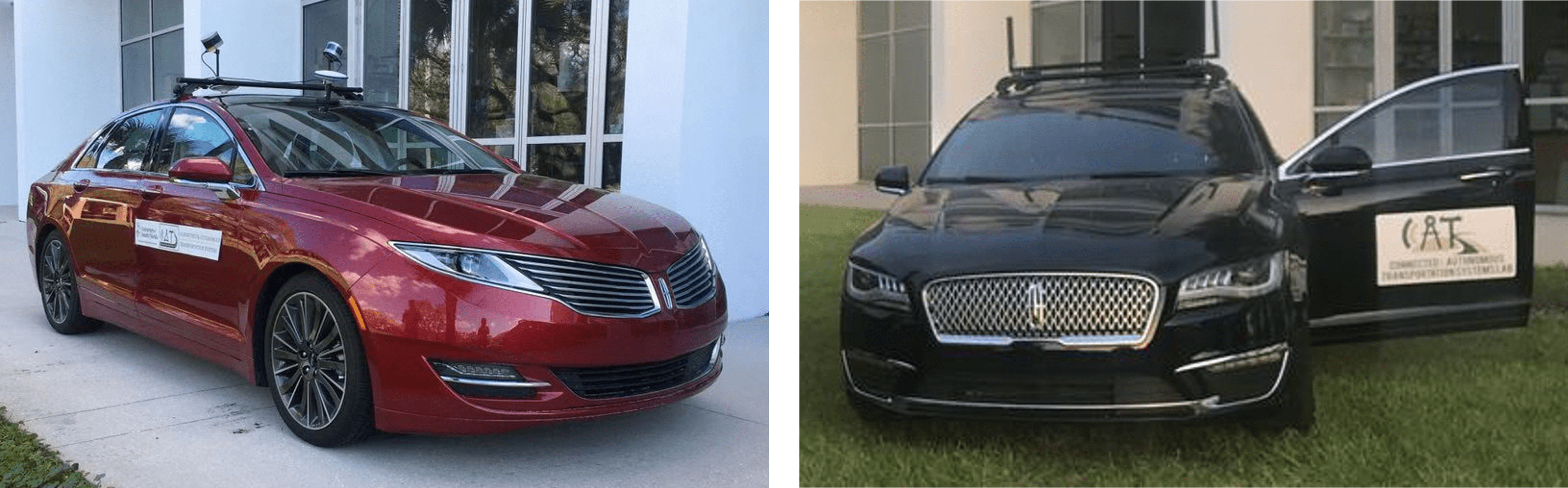}
\caption{AVs used to collected data in CATS dataset.}
\label{fig:AVs}
\end{figure}

The Connected and Autonomous Transportation Systems (CATS) Laboratory at University of Wisconsin-Madison has collected ADAS-equipped vehicle platooning dataset \cite{shi2021empirical}. The CATS ADAS-equipped vehicle platooning dataset documents commercial ADAS-equipped vehicle platooning behavior with different headway settings. Three vehicles participated in the experiments. The leading vehicle (i.e., ICE vehicle) operated with cruise control, following a predesigned speed profile. The following two vehicles (shown in Figure \ref{fig:AVs}) operated with ACC. The data collection occurred on a segment of State Road 56 in Florida between Bruce B. Downs Boulevard (Blvd.) and Gall Blvd. when traffic was light. The data collection team equipped each participating vehicle with a GPS device to record the real-time speed and location.


\subsection{Vanderbilt ACC Dataset}

Vanderbilt University collected Vanderbilt ACC data \cite{wang2019estimating}, which included car-following data of commercial ADAS-equipped vehicles with ACC enabled. In total, the data collection team recorded 1,200 miles of data. Participating vehicles were a lab-developed ADAS-equipped vehicle (hybrid vehicle) and commercial ADAS-equipped vehicles (ICE and hybrid vehicles) of two makes. The data collection team installed GPS devices on the vehicles to collect the vehicle data. This dataset includes two parts. The first part is the two-vehicle tests, in which the leading vehicle drove in a designed speed profile using cruise control, and the ACC vehicle followed. The data collection team used the same leading vehicle across different runs and tested different following vehicles. The second part is the eight-vehicle platoon tests. The lab ADAS-equipped vehicle served as the leading vehicle, following a precise speed profile. Seven other identical ACC vehicles followed.

\subsection{Central Ohio Dataset}

The Central Ohio dataset \cite{OhioOneVehicles,OhioTwoVehicles} captures naturalistic ADAS-equipped vehicle behavior on roadways in central Ohio. Two types of vehicles were used: readily identifiable ADAS-equipped vehicles and discreet ADAS-equipped vehicles. Both vehicles are Level 2 ADAS-equipped vehicles. The Ohio project collected and processed trajectories for adjacent vehicles in the traffic stream; thus, both vehicles are equipped with cameras and LiDAR to capture adjacent vehicle movements. For the readily identifiable ADAS-equipped vehicles, the radar and LiDAR sensors are visible on the vehicles. For the discreet ADAS-equipped vehicle, most sensors are not visible or were installed to be disguised (e.g., on top of the luggage stored on the top of the car). The team collected data from three generalized testing scenarios: vehicle following on highway and arterial roadways, lane-change on highway and arterial roadways, and intersection approach and departure. These three scenarios were evaluated in different traffic densities and involved either a single ADAS-equipped vehicle or multiple ADAS-equipped vehicles. The two types of data are published in \cite{OhioOneVehicles} and \cite{OhioTwoVehicles}, respectively. For the vehicle-following and lane-change scenarios on highways and freeways, the Ohio data collection team used Ohio's 33 Smart Mobility Corridor and Interstate 270 (I-270). The Ohio data collection team also collected data on urban arterial roadways, including Riverside Drive and Morse Road. 

\subsection{Waymo Open Dataset}

The Waymo Open Dataset offers high-resolution sensor data, amassed using ADAS-equipped vehicles across various environments, objects, and weather conditions \cite{sun2020scalability}. This dataset includes information from diverse locations such as San Francisco, Phoenix, Mountain View, Los Angeles, Detroit, and Seattle. It comprises two parts: a perception dataset and a motion dataset. As of the latest update in August 2023, the perception dataset, with high-resolution sensor data and labels, consists of 2030 segments, while the motion dataset contains object trajectories and corresponding 3D maps for 103354 segments. Each segment encompasses about 200 frames, with a 0.1-second interval between consecutive frames. However, it's important to note that these data do not directly include the relationships of car-following pairs.

Hu et al. \cite{hu2022processing} processed 1000 segments of the original Waymo Open Dataset released in 2019 and manually extracted car-following pairs from it. In the processed dataset, all crucial information pertinent to the driving behavior of AVs, surrounding vehicles, and other road users has been integrated into a single file. Rather than extract the car-following pairs through automatic methods, the authors manually extracted car-following pairs based on video analysis to ensure the quality and reliability of the data.

\section{METHODS}
\label{sec:method}

\subsection{The framework for data processing}

In our methodology, we implement a three-step process to restructure the datasets into a standardized, user-friendly format, specifically tailored for traffic flow research, with a particular focus on car following modeling in this paper. The first step involves the extraction of several critical features necessary for calibration, followed by the conversion of all datasets into a unified CSV format encompassing these features. The second step is dedicated to data cleaning, which involves the meticulous removal of noise, outliers, and extraneous information from the raw data. This is vital to ensure that the datasets accurately represent high-quality, real-world behavior. The final step encompasses the calibration of a linear model using these refined datasets, allowing us to conduct a comparative analysis of the results.

\subsection{Uniform Data Format}

The first process is to extract valuable features and transfer the data into a uniform format using CSV files. We aim to using the trajectory data to calibrate the car-following models. In this study, we focus on the linear car-following model, where the variables available as inputs to the upper-level controller include only the spacing between vehicles, relative velocity, and the velocity of the ACC vehicle itself. The linear model for the following vehicle is shown in Equation \eqref{equ:linear}.
\begin{align}
        \hat{a}(t)=f_s \times s(t)+f_v \times v(t)+f_{\Delta v} \times \Delta v(t)+z \label{equ:linear}
\end{align}
where $\hat{a}(t)$ is the predicted acceleration at time $t$, $s(t)$ is the spacing between the leading vehicle and the following vehicle at time $t$, $\Delta v(t)$ is the speed difference between the leading vehicle and the following vehicle at time $t$, $v(t)$ is the speed of the following vehicle at time $t$, $f_s, f_v, f_{\Delta v}, z$ are the parameters. Therefore, we obtain the main labels that are related to the model calibration in Table \ref{tab:labels}.

\begin{table}[h]
\caption{Selected features in the data.}
\begin{center}
        \begin{tabular}{lp{5cm}}
                \hline
                Labels & Description \\
                \hline
                traj\_id & The ID of one car-following trajectory pair. \\
                frame\_id & The ID of the frame. \\
                leader\_id & The ID of the leading vehicle. \\
                leader\_type & The vehicle type of the leading vehicle. 1 represents AV, 0 represents HV. \\
                leader\_speed & The speed of the leading vehicle. \\
                follower\_id & The ID of the following vehicle. \\
                follower\_type & The vehicle type of the following vehicle. 1 represents AV, 0 represents HV. \\
                follower\_speed & The speed of the following vehicle. \\
                follower\_acceleration & The acceleration of the following vehicle. \\
                spacing & The spacing between the leading vehicle and the following vehicle. \\
                speed\_diff & The speed difference of the leading vehicle and the following vehicle. \\
        \hline
        \end{tabular}
        \label{tab:labels}
\end{center}
\end{table}

In our approach to standardizing the GPS-based data, which typically includes vehicle speed and coordinates, we address the absence of spacing and acceleration data in some datasets. To achieve uniformity in data processing, we calculate the spacing by determining the distance between the leading and following vehicles, derived from their coordinates. For the computation of acceleration, we employ the first-order difference of vehicle speed. By standardizing spacing and acceleration calculations, the data becomes more reliably comparative, facilitating better analysis and interpretation. Finally, we consolidate all trajectory data from each dataset into a single CSV format file.

\subsection{Data Cleaning}

The second step is data cleaning. Prior to model calibration, we encountered various issues pertaining to files, data, and labels within the datasets. Consequently, it became imperative to undertake a thorough cleansing and pre-processing of the initial dataset. The data-cleaning process is specifically designed to address four key dimensions of challenges, ensuring the integrity and reliability of the data:

In light of the above issues plaguing the original dataset, we undertake a data cleaning and pre-processing regimen, delineated by the ensuing steps:
\begin{itemize}
        \item Step 1: Linear Interpolation for Missing Data. To address gaps in location data, linear interpolation techniques were employed to generate reasonable approximations. The interpolation step fills in data voids, thus augmenting the dataset's comprehensiveness and reducing the likelihood of errors in analyses dependent on location data.
        \item Step 2: Outlier Removal. All data points residing outside ±3 standard deviations from the mean were systematically eliminated to isolate abnormal data. The exclusion of outliers enhances the dataset's credibility, making it more amenable to statistical analysis.
        \item Step 3: Omission of Unstable Periods. Periods characterized by unstable car-following, specifically those recorded at the commencement or conclusion of test runs, were deliberately omitted. This exclusion preserves the quality and consistency of the dataset, thereby upholding its analytical reliability.
\end{itemize}

Through this comprehensive cleaning and pre-processing regimen, we substantially enhance the dataset's quality, aligning it more closely with the stringent requirements of automated vehicle technology research.

\subsection{Car-following Model Caliberation}

In this research, our primary focus is on the linear car-following model as delineated in \eqref{equ:linear}. The calibration of this model can be effectively resolved using the least squares method. However, it's important to recognize that the simplistic form of the linear model may not accurately capture the complexities of car-following behavior. In future work, we intend to extend our analysis to include non-linear models, such as the Intelligent Driver Model (IDM) \cite{treiber2000congested}, and compare their calibration results. This approach will allow us to explore the efficacy of various modeling techniques in replicating the nuances of car-following dynamics more accurately.

\section{EXPERIMENTS}
\label{sec:exp}

\subsection{Experiment Condition Setting}

In this study, we process three datasets: the AstaZero dataset in the OpenACC Dataset, the CATS Laboratory Open Dataset, and the Vanderbilt ACC Dataset. For each of these datasets, we implement the complete data processing method proposed in Section \ref{sec:method} and compare the statistical results of the processed data with the calibration data.

For these three datasets, we manually identified the range of unstable periods through observation. Notably, in the OpenACC dataset, when the vehicle speed approaches 0 (indicating the vehicle is still in the start-up phase and not yet engaging in car-following behavior), there is minor fluctuation in the spacing. To mitigate the impact of this data, we removed entries where the leading and following vehicle speeds were less than 0.1 $m/s$. Additionally, we excluded data where the acceleration was greater than +5 or less than -5 $m/s^2$.

In each dataset, we calibrate the AVs with all their related trajectories. In total, seven AVs were calibrated: an ACC-equipped SUV 2019 from the Vanderbilt ACC Dataset, a Lincoln MKZ 2016 and a Lincoln MKZ 2017 from the CATS Laboratory Open Dataset, and an Audi A6, a BMW X5, a Mercedes A-Class, and a Tesla Model 3 from the OpenACC Dataset. For the calibration of the OpenACC Dataset, a 2 seconds response time delay of the input variables is introduced as discussed in \cite{makridis2021openacc}. It is important to note that in the experimental section of this study, all distance data are in meters, speed data are in meters per second, and acceleration data are in meters per second squared.

\subsection{Statistical Results}

\begin{table*}[h]
        \caption{Statistical analysis results of the three datasets.}
        \label{table_example}
        \begin{center}
        \begin{tabular}{lcccccccccccc}
        \hline
        \multicolumn{1}{c}{\multirow{2}[0]{*}{Statistical measures}} & \multicolumn{4}{c}{Vanderbilt ACC Dataset} & \multicolumn{4}{c}{CATS Laboratory Open Dataset}      & \multicolumn{4}{c}{OpenACC Dataset} \\
                & $s$ & $v$ & $\Delta v$ & $a$ & $s$ & $v$ & $\Delta v$ & $a$ & $s$ & $v$ & $\Delta v$ & $a$ \\
        \hline
        max   & \multicolumn{1}{r}{56.35} & \multicolumn{1}{r}{34.34} & \multicolumn{1}{r}{2.07} & \multicolumn{1}{r}{0.57} & \multicolumn{1}{r}{61.04} & \multicolumn{1}{r}{25.41} & \multicolumn{1}{r}{1.95} & \multicolumn{1}{r}{0.79} & \multicolumn{1}{r}{62.15} & \multicolumn{1}{r}{33.16} & \multicolumn{1}{r}{3.00} & \multicolumn{1}{r}{2.41} \\
        min   & \multicolumn{1}{r}{16.95} & \multicolumn{1}{r}{25.67} & \multicolumn{1}{r}{-2.52} & \multicolumn{1}{r}{-0.62} & \multicolumn{1}{r}{20.71} & \multicolumn{1}{r}{20.58} & \multicolumn{1}{r}{-1.96} & \multicolumn{1}{r}{-0.80} & \multicolumn{1}{r}{5.18} & \multicolumn{1}{r}{3.47} & \multicolumn{1}{r}{-3.00} & \multicolumn{1}{r}{-2.42} \\
        mean  & \multicolumn{1}{r}{35.85} & \multicolumn{1}{r}{29.23} & \multicolumn{1}{r}{0.02} & \multicolumn{1}{r}{0.01} & \multicolumn{1}{r}{42.17} & \multicolumn{1}{r}{23.23} & \multicolumn{1}{r}{-0.01} & \multicolumn{1}{r}{-0.01} & \multicolumn{1}{r}{24.56} & \multicolumn{1}{r}{18.49} & \multicolumn{1}{r}{0.02} & \multicolumn{1}{r}{0.01} \\
        std   & \multicolumn{1}{r}{6.19} & \multicolumn{1}{r}{1.59} & \multicolumn{1}{r}{0.78} & \multicolumn{1}{r}{0.19} & \multicolumn{1}{r}{10.41} & \multicolumn{1}{r}{0.83} & \multicolumn{1}{r}{0.70} & \multicolumn{1}{r}{0.24} & \multicolumn{1}{r}{9.53} & \multicolumn{1}{r}{4.17} & \multicolumn{1}{r}{0.84} & \multicolumn{1}{r}{0.49} \\
        \#sample & \multicolumn{4}{c}{8752}      & \multicolumn{4}{c}{3577}      & \multicolumn{4}{c}{401411} \\
        \#trajectory & \multicolumn{4}{c}{1 trajectory for each vehicle} & \multicolumn{4}{c}{7 trajectories for each vehicle} & \multicolumn{4}{c}{8 trajectories for each vehicle} \\
        \hline
        \end{tabular}%
        \label{tab:stat}
\end{center}
\end{table*}

\begin{figure*}[ht]
\centering
\includegraphics[width=1.0\linewidth]{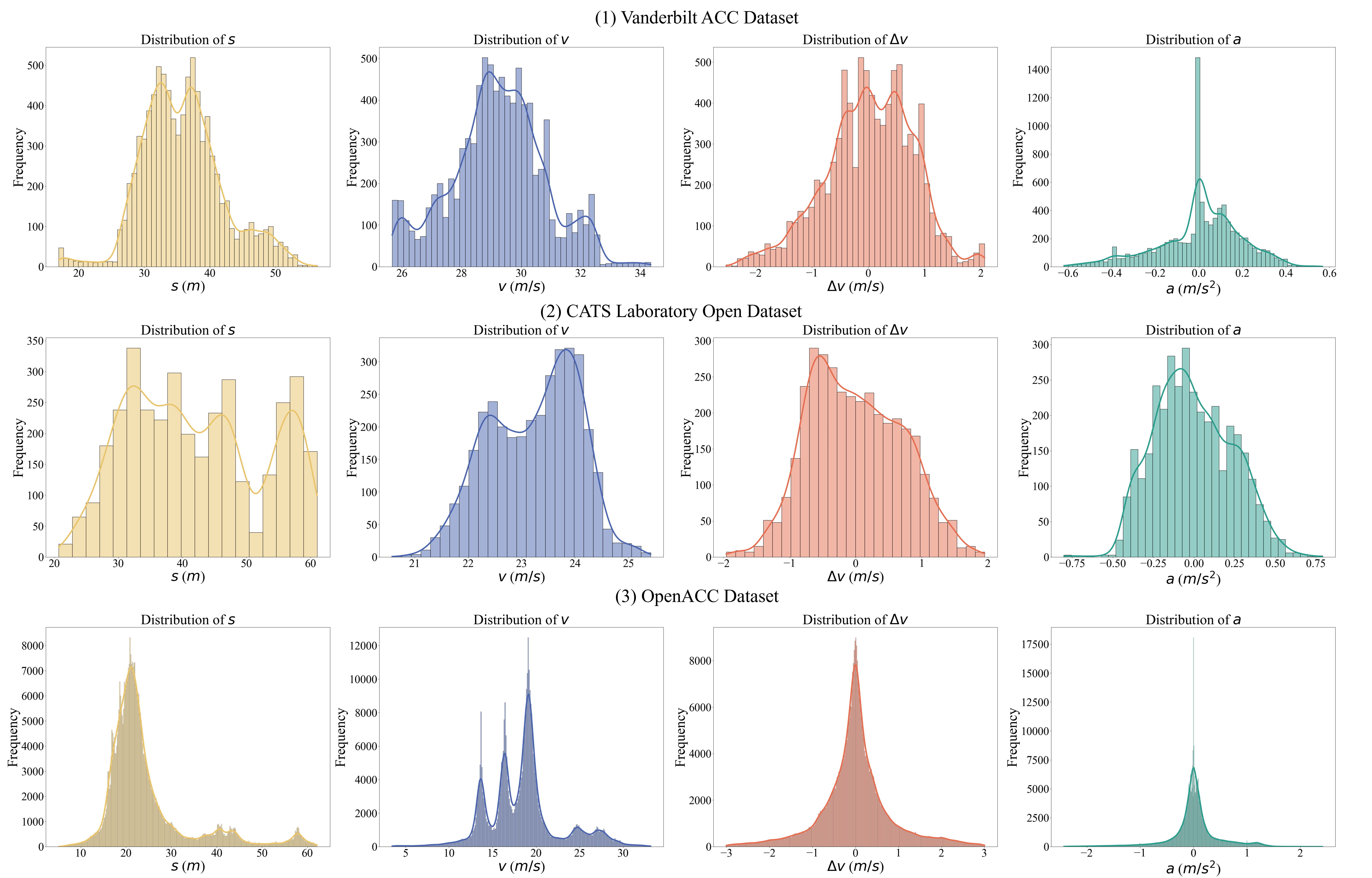}
\caption{Distributions of the main features for the three datasets.}
\label{fig:distribution}
\end{figure*}

The statistical analysis and the distribution of the main features for the processed datasets are detailed in Table \ref{tab:stat} and Figure \ref{fig:distribution}, where the notations are the same with Section \ref{sec:method} and $a$ represents the acceleration of the following vehicle. The results reveal that the speed difference and the following vehicle's acceleration across the three datasets are relatively consistent, predominantly ranging from -2 $m/s$ to 2 $m/s$ and -1 $m/s^2$ to 1 $m/s^2$, respectively, with mean values near 0. However, significant disparities are observed in terms of spacing and the following vehicle's speed. For instance, the OpenACC Dataset exhibits spacing evenly distributed between 15m and 30m, averaging 24.56m, and a following vehicle's speed primarily between 12 $m/s$ to 20 $m/s$, averaging 18.49 $m/s$. This distribution suggests an environment characterized by relatively small spacing and lower speeds. Conversely, the Vanderbilt ACC Dataset predominantly features spacing between 26m and 50m, averaging 35.85 $m$, with the following vehicle's speed mostly in the 25 $m/s$ to 32 $m/s$ range, averaging $29.23 m/s$. The CATS Laboratory Open Dataset shows a uniform spacing distribution between 25 $m$ and 60 $m$, averaging 42.17 $m$, and the following vehicle's speed concentrated between 21m/s and 25 $m/s$, averaging 23.23 $m/s$. Both the Vanderbilt ACC Dataset and CATS Laboratory Open Dataset indicate environments with moderate spacing and speed. These findings suggest that models calibrated from the OpenACC dataset are better suited for low-speed ranges, while those from the Vanderbilt ACC Dataset and CATS Laboratory Open Dataset are more applicable to medium-speed ranges. Moreover, in terms of dataset size, the OpenACC dataset, with over 400 thousand samples, is substantially larger than the other two datasets, each comprising less than 10 thousand samples.

\subsection{Calibration Results}

The calibration results of the linear models are presented in Table \ref{tab:calibration}. Here we use the $R^2$ to measure the accuracy of the linear model. From Table \ref{tab:calibration}, it can be observed that the $R^2$ for the linear models of most vehicles range between 0.5 and 0.7, except the two CATS vehicles that are much higher than 0.7. This indicates that the linear models can reflect AV car-following behavior to a certain extent. However, due to the inherent limitations of the model structure and the complexity of the ACC systems, linear models struggle to perfectly simulate car-following behavior. Analyzing from the perspective of model parameters, parameters $f_d$ and $f_{\Delta v}$ for all vehicles are positive. This implies that a larger spacing and speed difference lead to a greater acceleration of the following vehicle. However, regarding parameter $f_v$, the values for the Vanderbilt SUV and OpenACC AVs are negative, whereas only the CATS AVs' parameter is positive. Moreover, the intercept term $z$ in the CATS AVs model is very small, suggesting that the model predicts a negative acceleration value at the initial phase of the vehicles' start. Considering the range of the datasets' distribution, we infer that the linear model derived from the CATS dataset may only be valid in scenarios where the speed is predominantly above zero and is not applicable for low-speed conditions. Another possible reason for this discrepancy is that the CATS data was collected at a frequency of 1 Hz, which is a large time interval that may not accurately calibrate the model.

\begin{table}[htbp]
        \centering
        \caption{Linear model parameters for the test vehicles.}
        \begin{tabular}{lcccccc}
        \hline
        Vehicle & $R^2$ & $f_s$ & $f_v$ & $f_{\Delta v}$ & $z$ \\
        \hline
        Vanderbilt SUV & 0.6561 & 0.0165 & -0.0067 & 0.1532 & -0.3921 \\
        Lincoln MKZ 2016 & 0.7422 & 0.0009 & 0.1733 & 0.3953 & -4.0682 \\
        Lincoln MKZ 2017 & 0.8198 & 0.0012 & 0.1940 & 0.4022 & -4.5568 \\
        Audi A6 & 0.6318 & 0.0038 & -0.0010 & 0.4346 & -0.0645 \\
        BMW X5 & 0.6318 & 0.0061 & -0.0014 & 0.4838 & -0.1214 \\
        Mercedes AClass & 0.6556 & 0.0057 & -0.0049 & 0.3910 & -0.0447 \\
        Tesla Model3 & 0.7075 & 0.0036 & -0.0019 & 0.5767 & -0.0566 \\
        \hline
        \end{tabular}%
        \label{tab:calibration}%
\end{table}%

\section{CONCLUSIONS}
\label{sec:con}

In this paper, we conducted an extensive review of five major trajectory datasets in the literature relevant to ADAS, focusing our analysis on three of them. Our study began with a detailed exposition of these datasets, encompassing their key features, data volume, formatting, and the environments in which the tests were conducted. We proposed a structured data processing methodology, which involves three critical steps: standardizing the data into a uniform format, data cleaning, and car-following model calibration. The experimental section of the paper delves into the statistical and calibration results. Our findings indicate that the CATS and Vanderbilt ACC datasets are more apt for models operating at medium speeds, whereas the OpenACC dataset is better suited for low-speed models. Future directions for this research include extending our data processing method to the remaining datasets. Additionally, we envisage undertaking a review and data processing of some trajectory datasets with short segments of less than 15 seconds that are designed for motion prediction.

\bibliographystyle{IEEEtran}
\bibliography{reference}

\end{document}